\documentstyle[twoside,graphicx]{elsart}
\graphicspath{{Figures/}{.}}
\begin{document} 
\begin{frontmatter}
\title{\bf Order Parameter Equations for Front Transitions: Nonuniformly
Curved Fronts}  
\author{Aric Hagberg\thanksref{ARIC}}
\address{Center for Nonlinear Studies and T-7, Theoretical Division,\\
Los Alamos National Laboratory, Los Alamos, NM 87545}
\thanks[ARIC]{\tt aric@lanl.gov}

\author{Ehud Meron\thanksref{EHUD}}
\address{The Jacob Blaustein Institute for Desert Research and the Physics 
Department,\\ Ben-Gurion University, Sede Boker Campus 84990, Israel}
\thanks[EHUD]{\tt ehud@bgumail.bgu.ac.il}

\date{\today}

\begin{abstract}
Kinematic equations for the motion of slowly propagating, weakly
curved fronts in bistable media are derived. The equations generalize
earlier derivations where algebraic relations between the normal front
velocity and its curvature are assumed. Such relations do not capture
the dynamics near nonequilibrium Ising-Bloch  (NIB) bifurcations, where
transitions between counterpropagating Bloch fronts may
spontaneously occur. The kinematic equations consist of coupled 
integro-differential equations for the front curvature and the front
velocity, the order parameter associated with the NIB bifurcation.
They capture the
NIB bifurcation, the instabilities of Ising and Bloch fronts to
transverse perturbations, the core structure of a spiral wave, and the
dynamic process of spiral wave nucleation.
\end{abstract}
\end{frontmatter}

%
%
\section{Introduction} 
Interfaces separating different equilibrium or nonequilibrium states
appear in a variety of contexts including crystal growth, domain walls
in magnetic and hydrodynamic systems, and reaction-diffusion
fronts~\cite{CrHo:93}. 
The global patterns that appear in these systems depend to a
large extent on the possible occurrence of interfacial
instabilities. A transverse instability of the interface, 
for example, may lead to fingering and the formation of
labyrinthine 
patterns~\cite{Haslam:88,LMOS:93,Pear:93,LeSw:95,HaMe:94c,GMP:96,MuOS:96}.
Another
instability with dramatic effects on pattern formation is the
nonequilibrium Ising-Bloch (NIB)
bifurcation~\cite{RiTe:82,IMN:89,CLHL:90,BRSP:94,HaMe:94a}. 
The bifurcation, which
takes a single stable (Ising) front to a pair of counterpropagating
stable (Bloch) fronts, has been found in chemical
reactions~\cite{HBKR:95,LOS:96,Haim:96} and in liquid
crystals~\cite{FRCG:94,Frisch:94,NYK:95}.  Far below the NIB
bifurcation stationary patterns or uniform states prevail. Far beyond
it, a regime of ordered traveling patterns, including spiral waves,
exists. In the
vicinity of the bifurcation disordered spatio-temporal patterns,
involving repeated events of spiral-wave nucleation appear. This
behavior, which we call ``spiral turbulence'', 
has been attributed to spontaneous transitions
between counterpropagating Bloch fronts induced by curvature
variations, front interactions, and interactions with 
boundaries~\cite{HaMe:94c,EHM:95}.

A common theoretical approach to studying pattern formation in
interfacial systems is based on a geometric equation for the 
interface curvature (see Eqn.~(\ref{K})
below)~\cite{BKKL:84,Zykov:87,Mikhailov:90,Meron:92,MDZy:94}.  Once the
dependence of the normal velocity of the interface on its curvature is
known the shape of the interface can be determined at any given
time. For reaction-diffusion fronts this dependence may become
particularly simple: Away from a NIB bifurcation, a linear relation is
an excellent approximation~\cite{Meron:92,TyKe:88,FMH:88}. 
The curvature equation, however, does not
capture possible transitions between counterpropagating fronts near a
NIB bifurcation because the front velocity becomes an independent
slow degree of freedom and can no longer 
algebraically be related to curvature~\cite{EHM:95,HMRZ:97,Bode:96}.

In this paper we consider bistable media that exhibit NIB bifurcations and 
derive kinematic front equations which generalize the geometric curvature
equation. The new kinematic equations capture transitions between
counterpropagating fronts, and  spontaneous spiral-wave 
nucleation, a process which plays a crucial role in the
onset of spiral turbulence. 
The equations are:
\begin{itemize}
\item An equation for the order parameter, $C_0$, associated with the NIB 
bifurcation:
\begin{equation}
{\partial C_0\over\partial t}=(\alpha_c-\alpha)C_0 - \beta C_0^3
+\gamma\kappa + \gamma_0 +{\partial^2 C_0\over \partial s^2} 
- {\partial C_0 \over \partial s} \int_0^s \kappa C_n ds^\prime\,. 
\label{C0}
\end{equation}
\item A geometric equation for the front curvature, $\kappa$:
\begin{equation}
{\partial\kappa\over\partial t} = -(\kappa^2 + {\partial^2\over\partial
s^2})C_n - {\partial\kappa\over\partial s}\int_0^s \kappa C_n ds^\prime \,.
\label{K}
\end{equation}
\item An equation  relating the normal front velocity $C_n$, the curvature
$\kappa$, and the order parameter, $C_0$:
\begin{equation}
C_n = C_0 - D\kappa\,.
\label{Cn}
\end{equation}
\end{itemize}
In these equations $s$ is the front arclength, and the critical
parameter value, $\alpha_c$ designates the NIB bifurcation point. Note that
Eqn. (\ref{Cn}) cannot be regarded as a linear relation between the normal
velocity of the front and its curvature since 
$C_0$ is not a constant but a
dynamical variable coupled to curvature through Eqn.~(\ref{C0}).
In fact, Eqns. (\ref{C0}) and (\ref{Cn}) can be recast into a single
integro-differential equation for the normal velocity (using
Eqn. (\ref{K}))
\begin{equation}
{\partial C_n\over\partial t} = {\cal  F}
\left[C_n,\kappa;\frac{\partial}{\partial s}\right] - 
{\partial C_n\over\partial s}\int_0^s \kappa C_n ds^\prime \,,
\label{Cndif}
\end{equation}
which replaces the algebraic $C_n-\kappa$ relation used in earlier
derivations. 
An algebraic $C_n-\kappa$ relation can be recovered from
Eqns.~(\ref{C0}) and (\ref{Cn}) assuming the order parameter
$C_0$ follows adiabatically slow curvature variations.
This issue is further discussed at the end of Section~3. 

The order parameter equation (\ref{C0}) yields the NIB bifurcation for
planar (uncurved) fronts. For a symmetric bistable system
($\gamma_0=0$), the Ising front, $C_0=0$, is stable for
$\alpha>\alpha_c$. At $\alpha=\alpha_c$ the Ising front becomes
unstable and a pair of Bloch fronts appears,
$C_0=C_0^\pm\equiv\pm\sqrt{(\alpha_c-\alpha)/\beta}$. For a
nonsymmetric system ($\gamma_0\ne 0$) this pitchfork bifurcation
unfolds into a saddle node bifurcation in the usual way.

A brief account of the results to be reported here has appeared in
Ref.~\cite{HaMe:97}. We present in Section 3 a detailed derivation of
the kinematic equations for a particular reaction-diffusion model
introduced in Section 2. In Section 4 we study the kinematic
equations. We analyze the stability of planar fronts to transverse 
perturbations and present numerical solutions describing steadily
rotating spiral waves and spiral-wave nucleation induced by a
transverse instability.
We conclude in Section 5 with a discussion of our results.

%
%

\section{The reaction diffusion model}

We consider the FitzHugh-Nagumo model with a diffusing inhibitor,
\begin{eqnarray}
{\partial u\over\partial t}&=& 
\epsilon^{-1}(u-u^3-v)+\delta^{-1}\nabla^2u\,,\nonumber \\
{\partial v\over\partial t}&=& u-a_1v-a_0+\nabla^2 v\,, \label{FHN}
\end{eqnarray} 
where $u$ and $v$, the activator and the inhibitor, are real scalar fields
and $\nabla^2$ is the Laplacian operator in two dimensions. The parameter
$a_1$ is chosen so that~(\ref{FHN}) describes a bistable medium having two
stable uniform states: an up state $(u_+,v_+)$ and a down state
$(u_-,v_-)$. 
Ising and Bloch front solutions connect the two uniform states
$(u_\pm,v_\pm)$ as the spatial coordinate normal to the front goes from 
$-\infty$ to $+\infty$.
The parameter space of interest is spanned by $\epsilon,
\delta$ and $a_0$, or alternatively by 
$\eta=\sqrt{\epsilon\delta}$, $\mu=\epsilon/\delta$,
and $a_0$. Note the parity symmetry
$(u,v)\to(-u,-v)$ of~(\ref{FHN}) for $a_0=0$. 

\begin{figure}
\centering\includegraphics[width=5.0in]{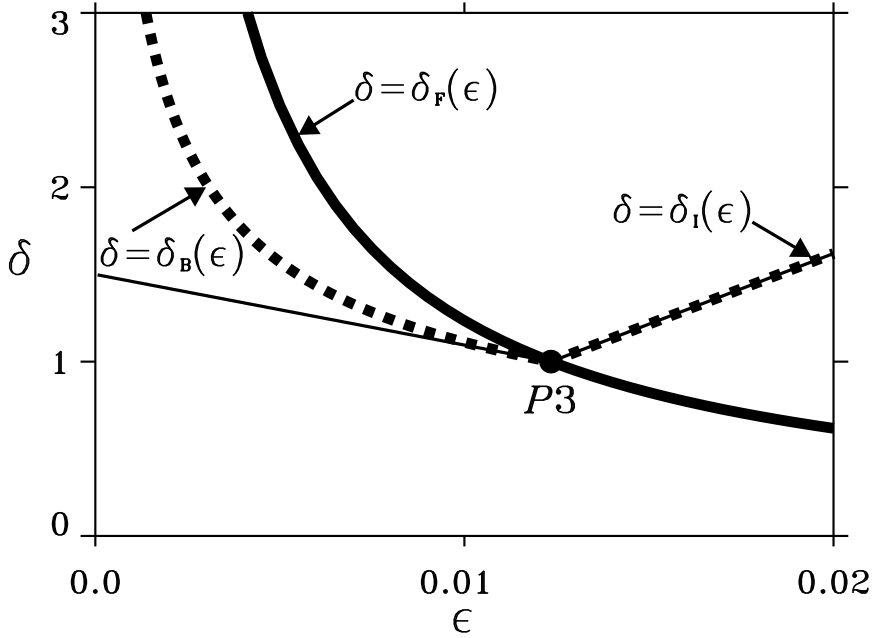}
\caption{
The NIB front bifurcation and planar front transverse instability
    boundaries in the $\epsilon-\delta$ parameter plane.  The thick
    line is the NIB bifurcation,
    $\delta_F(\epsilon)=\eta_c^2/\epsilon$.  
    The dashed
    lines are the boundaries for the transverse instability of Ising,
    $\delta_I(\epsilon)$, and Bloch, $\delta_B(\epsilon)$, fronts.
    When $\delta>\delta_I$ ($\delta>\delta_B$) planar Ising (Bloch)
    fronts are unstable to transverse perturbations.
    The thin lines are approximations to the
    transverse instability boundaries obtained from a linear
    stability analysis of the kinematic
    equations.  Parameters: $a_1=4.0$, $a_0=0$.
}
\label{fig:lines}
\end{figure}

The NIB bifurcation line for $a_0=0$ is shown in
Fig.~\ref{fig:lines}. For $\mu\ll 1$ it is given by
$\delta=\delta_F(\epsilon)=\eta_c^2/\epsilon$, or $\eta=\eta_c$, where
$\eta_c=\frac{3}{2\sqrt{2}q^3}$ and $q^2=a_1+1/2$~\cite{HaMe:94a}.
The single stationary Ising front that exists for $\eta>\eta_c$ loses
stability to a pair of counterpropagating Bloch fronts at
$\eta=\eta_c$.  Beyond the bifurcation ($\eta<\eta_c$) a Bloch front
pertaining to an up state invading a down state coexists with another
Bloch front pertaining to a down state invading an up state. Also
shown in Fig.~\ref{fig:lines} are the transverse instability
boundaries (for $a_0=0$), $\delta=\delta_I(\epsilon)=
\epsilon/\eta_c^2$ and $\delta=\delta_B(\epsilon)=\eta_c/\sqrt\epsilon$,
for Ising and Bloch fronts respectively.  Above these lines,
$\delta>\delta_{I,B}$, planar fronts are unstable to transverse
perturbations~\cite{HaMe:94c,HaMe:94b}.  All three lines meet at a
codimension 3 point $P3$: $\epsilon=\eta_c^2$, $\delta=1$, $a_0=0$.

\section{Deriving the kinematic equations} 
The derivation of the kinematic equations is described here in three 
steps: A transformation to a frame moving with the curved front (Section 3.1), 
derivation of the normal velocity equation (\ref{Cn}) using a singular
perturbation approach (Section 3.2), and derivation of the order parameter
equation (\ref{C0}) (Section 3.3). In deriving the equations we assume that
$\mu=\epsilon/\delta\ll 1$ and that curvature is small, $\kappa\ll 1$.
Additional assumptions needed in deriving the order parameter equation
are described in Section 3.3.

\subsection{The moving frame}

We transform to an orthogonal coordinate
system $(r,s)$ that moves with the front, where $r$ is a
coordinate normal to the front and $s$ is the arclength.  
We denote the position vector of the front by  ${\bf X}(s,t)=(X,Y)$, and let it
coincide with the $u=0$ contour line. The unit vectors tangent and normal to
the front are given by
\[
{\bf \hat s}=\cos\theta{\bf \hat x} + \sin\theta{\bf \hat y}\,, \qquad
{\bf \hat r}=-\sin\theta{\bf \hat x} + \cos\theta{\bf \hat y}\,,
\]
where $\theta(s,t)$ is the angle that $\hat s$ makes with the $x$ axis.
A point ${\bf x}=(x,y)$ in the laboratory frame can be expressed as
\[
{\bf x}={\bf X}(s,t)+r{\bf \hat r}\,.
\]
This gives the following relation between the laboratory coordinates $(x,y)$ 
and the coordinates $(s,r)$ in the moving frame:
\begin{equation}
x=X(s,t)-r{\partial Y\over\partial s}\,,  \qquad  
y=Y(s,t)+r{\partial X\over\partial s}\,,
\label{trans}
\end{equation}
where we used the fact that ${\bf \hat s}=\partial{\bf X}/\partial s$.

With this coordinate change, partial spatial derivatives transform according to
\begin{equation}
\frac{\partial}{\partial x}=-\frac{\partial Y}{\partial s}
\frac{\partial}{\partial r}+
G\frac{\partial X}{\partial s}\frac{\partial}{\partial s}\,, \qquad
\frac{\partial}{\partial y}=\frac{\partial X}{\partial s}
\frac{\partial}{\partial r}+
G\frac{\partial Y}{\partial s}\frac{\partial}{\partial s} \,,
\label{partder}
\end{equation}
where  
\[ 
G=(1+r\kappa)^{-1}\,,
\]
and $\kappa$, the front curvature, is given by
\[
\kappa=-\frac{\partial\theta}{\partial s}=\frac{\partial Y}{\partial s}
\frac{\partial^2 X}{\partial s ^2}-\frac{\partial X}{\partial s}
\frac{\partial^2 Y}{\partial s ^2}\,.
\]
The time derivative transforms according to
\begin{equation}
{\cal D}=\frac{\partial}{\partial t}+
\frac{\partial r}{\partial t}\frac{\partial}{\partial r}+
\frac{\partial s}{\partial t}\frac{\partial}{\partial s}\,,
\label{D}
\end{equation}
where ~\cite{Mikhailov:90,Meron:92}
\begin{eqnarray}
{\partial s\over\partial t}&=&
\int_0^s \kappa C_n ds^\prime \,,\label{st}\\
{\partial r\over\partial t}&=&-C_n\,.
\end{eqnarray}
Using (\ref{partder}) we find for the Laplacian operator
frame~\cite{Keener:80}
\begin{equation}
\nabla^2={\partial^2\over\partial r^2} + \kappa G{\partial\over\partial r}
+ G{\partial G\over\partial s}{\partial\over\partial s} +
G^2{\partial^2\over\partial s^2}\,.
\label{nabla}
\end{equation}
The reaction-diffusion system (\ref{FHN}) in the moving frame is
\begin{eqnarray}
\eta\sqrt{\mu}{\cal D}u&=& 
u-u^3-v + \mu\nabla^2 u\,,\nonumber \\
{\cal D}v&=& u-a_1v-a_0+ \nabla^2 v\,, \label{mFHN}
\end{eqnarray}
where ${\cal D}$ and $\nabla^2$ are given by (\ref{D}) and (\ref{nabla}),
respectively, and we recall that $\mu=\epsilon/\delta$ and 
$\eta=\sqrt{\epsilon\delta}$.

\subsection{The normal velocity equation}
We study Eqns.~(\ref{mFHN}) assuming $\mu\ll
1$. Note that the limit $\mu\to 0$ can be taken safely without departing from
the immediate neighborhood of the front bifurcation at
$\eta=\eta_c$. We will use this fact in deriving the order parameter equation.
Consider the narrow front region where $u$ changes on a spatial scale of
order $\sqrt{\mu}$ while $v$ changes on a scale of order unity. Stretching the
normal coordinate $r$ according to $z=r/\sqrt{\mu}$, Eqns.~(\ref{mFHN})
become 
\begin{eqnarray}
u-u^3-v+\frac{\partial^2 u}{\partial z^2}+
\sqrt{\mu}\big(-\eta{\cal D}u +\kappa G\frac{\partial u}{\partial z}\bigr)+
\mu\bigl(G^2\frac{\partial^2 u}{\partial s^2}+G\frac{\partial G}{\partial s}
\frac{\partial u}{\partial s}\bigr)&=&0 \,,\nonumber \\
\frac{\partial^2 v}{\partial z^2}-\mu\bigl({\cal D}v+u-a_1v-a_0
-G^2\frac{\partial^2 v}{\partial s^2}-G\frac{\partial G}{\partial s}
\frac{\partial v}{\partial s}\bigr)&=&0\,. \label{inner}
\end{eqnarray}
Expanding $u$ and $v$ as
\begin{eqnarray}
u&=&u_0+\sqrt{\mu} u_1 + \mu u_2 +... \,,\nonumber \\
v&=&v_0+\sqrt{\mu} v_1 + \mu v_2 +... \,,\nonumber
\end{eqnarray}
we find at order unity the stationary front solution
\[ u_0=-\tanh(z/\sqrt 2),\qquad v_0=0\,. \]
At order $\sqrt{\mu}$ we find the equations
\begin{equation}
{\cal L}u_1=v_1+\eta\frac{\partial z}{\partial t}\frac{\partial u_0}
{\partial z}-\kappa G\frac{\partial u_0}{\partial z}\,, \qquad\qquad
\frac{\partial^2 v_1}{\partial z^2}=0\,, \label{order1}
\end{equation}
where
\begin{equation}
{\cal L}=\frac{\partial^2 }{\partial z^2}+1-3u_0^2\,.
\end{equation}
Solvability of (\ref{order1}) yields
\begin{equation}
{\partial r\over\partial t}={3\over\eta\sqrt 2}v_f +
\delta^{-1}\kappa\ ,
\label{rt}
\end{equation}
where $v_f=v(0,s,t)+{\cal O}(\epsilon^2)$ is the approximately
constant value of the inhibitor $v$ in the narrow [${\cal
O}(\sqrt\mu)$] front core region. 
The first term on the
right-hand-side of~(\ref{rt}) is identified with the order parameter
for the NIB bifurcation: $C_0=-{3\over\eta\sqrt 2}v_f$.  Since the
normal velocity is $C_n=-{\partial r\over\partial t}$, Eq.~(\ref{rt})
yields the normal velocity relation~(\ref{Cn}) with $D=\delta^{-1}$.

\subsection{The order parameter equation}

Away from the narrow front region $u$ and $v$ change on the same spatial
scales. Letting $\mu\to 0$ in Eqs. (\ref{mFHN}) we obtain $u-u^3-v=0$. 
The relevant solutions
are $u=u_+(v)\approx 1-v/2$ for $r<0$ and $u=u_-(v)\approx -1-v/2$ for
$r>0$ (assuming $a_1$ is sufficiently large)~\cite{HaMe:94a}. We thus obtain 
the following free boundary problem for $v$:
\begin{eqnarray}
{\cal M}v
&=&+1 -{3\over \eta\sqrt 2}v_f{\partial v\over\partial r}+P_1+P_2,
\quad r\le 0\,, \nonumber \\
{\cal M}v
&=&-1 -{3\over \eta\sqrt 2}v_f{\partial v\over\partial r}+P_1+P_2,
\quad r\ge 0\,, \nonumber \\
&&v(\mp\infty,s,t)=v_\pm={\pm 1 - a_0\over q^2}\ ,\nonumber \\
&&\left[v\right]_{r=0}=\left[{\partial v\over\partial r}\right]_{r=0}=0\ ,
\label{free}
\end{eqnarray}
where ${\cal M} = {\partial \over\partial t}-{\partial^2\over\partial r^2}+q^2,$
\begin{eqnarray*}
P_1&=&\left(1-\delta^{-1}\right)\kappa{\partial v\over\partial r}-a_0+
G^2{\partial^2
v\over \partial s^2}-{\partial s\over\partial t}
{\partial v\over\partial s}\,, 
\\
P_2&=&G{\partial G\over\partial s}{\partial v\over\partial s}\, ,
\end{eqnarray*}
and the square brackets denote jumps of the quantities inside the brackets
across the front at $r=0$.

We confine ourselves to nearly symmetric systems ($\vert a_0\vert\ll 1$)
and to a parameter regime that includes the immediate vicinity of $P3$ 
and extends into the Ising regime near or below the transverse instability 
boundary $\delta=\delta_I(\epsilon)$.
This allows solving the free boundary problem~(\ref{free}) 
by expanding propagating curved front solutions as power
series in $c$ around the stationary planar Ising front~\cite{HMRZ:97,HMRZ:96}, 
where $c\ll 1$ is the speed of a planar Bloch front solution. We  
assume weak dependence of $\kappa$ and $v$ on $s$ and achieve this
by introducing the slow length scale $S=cs$ and assuming ${\bf X}=
{\bf X}(S,t)$.  This assumption dictates $\kappa=c^3\kappa_0$  where
$\kappa_0\sim {\cal O}(1)$. We also introduce a slow time scale 
$T=c^2t$ to describe deviations from steady front motion. We write
\begin{equation}
v(r,S,t,T)=v^{(0)}(r)+\sum_{n=1}^\infty c^nv^{(n)}(r,S,t,T)\ ,
\label{vrStT}
\end{equation}
where 
\begin{eqnarray*}
v^{(0)}(r)&=q^{-2}(1-e^{qr})\,,& \qquad r\le 0\,, \\
v^{(0)}(r)&=q^{-2}(e^{-qr}-1)\,,& \qquad r\ge 0\,. \nonumber
\end{eqnarray*}
Expanding 
\[
\eta=\eta_c-c^2\eta_1+c^4\eta_2+... \,,
\]
anticipating a pitchfork bifurcation, and 
using these expansions in 
(\ref{free}) produces the set of equations
\begin{equation}
\frac{\partial v^{(n)}}{\partial t} + q^2 v^{(n)} - 
\frac{\partial^2 v^{(n)}}{\partial r^2} =
 -\rho^{(n)},~n=1,2,3,\ldots
\label{vn}
\end{equation}
where 
\begin{eqnarray}
\rho^{(1)}&=&
{3\over\sqrt2\eta_c}v^{(1)}_{\vert r=0}
\frac{\partial v^{(0)}}{\partial  r}\,,\\
\rho^{(2)}&=&
{3\over\sqrt2\eta_c}
\left[v^{(1)}_{\vert r=0}\frac{\partial v^{(1)}}{\partial r}+
v^{(2)}_{\vert r=0}\frac{\partial v^{(0)}}{\partial r}\right]\,, \\  
\rho^{(3)}&=&
+{3\eta_1\over\sqrt 
2\eta_c^2}v^{(1)}_{\vert r=0}\frac{\partial v^{(0)}}{\partial r} 
+{3\over\sqrt2\eta_c}
\left[v^{(1)}_{\vert r=0}\frac{\partial v^{(2)}}{\partial r}
+v^{(2)}_{\vert r=0}\frac{\partial v^{(1)}}{\partial r} 
+v^{(3)}_{\vert r=0}\frac{\partial v^{(0)}}{\partial r}\right]  \label{rho3}\\
&&\mbox{}
+V(r,S,T)+a_{00}-(1-\delta^{-1})\kappa_0 \frac{\partial
v^{(0)}}{\partial r} \,,\nonumber
\end{eqnarray}
and
\begin{equation}
V(r,S,T)=\frac{\partial v^{(1)}}{\partial T}-G^2 \frac{\partial^2 v^{(1)}}
{\partial S^2}+\frac{\partial S}{\partial T}\frac{\partial v^{(1)}}{\partial S} \,.
\label{vrst}
\end{equation}
In (\ref{rho3}) we assumed $a_0=c^3 a_{00}$ where $a_{00}\sim {\cal 
O}(1)$, and recall that $\kappa_0=\kappa/c^3$. Notice that ${\partial
S\over\partial T}\sim{\cal O}(1)$, and $P_2$ contributes only at orders 
higher than $c^3$. 

Equations (\ref{vn}) should be supplemented by appropriate asymptotic
conditions as $r\to\pm\infty$. Since $\lim_{r\to\mp\infty}v^{(0)}(r)=
\pm q^{-2}$ and $a_0\sim {\cal O}(c^3)$, the asymptotic conditions 
in (\ref{free}) are satisfied by demanding
\begin{equation}
\lim_{r\to\pm\infty}v^{(n)}=0 \qquad n=1,2 \,,
\label{as_vn}
\end{equation}
and
\begin{equation}
\lim_{r\to\pm\infty}v^{(3)}=-a_{00}/q^2 \,.
\label{as_v3}
\end{equation}
We are interested in solutions of Eqns~(\ref{vn}) at long times where 
they become independent of the fast
time scale $t$: $\partial v^{(n)}/\partial t \to 0$ as $t\to\infty$. 
For $n=1$, the stationary 
solution of (\ref{vn}) with the asymptotic condition (\ref{as_vn}) 
is~\cite{HMRZ:97}
\[
v^{(1)}(r,S,T)=v^{(1)}_{\vert r=0}F(r)\,,\label{fun1}
\]
where
\begin{eqnarray}
F(r)&=(1-qr)e^{qr}\,,& \qquad r\le 0\,, \nonumber \\
F(r)&=(1+qr)e^{-qr}\,,& \qquad r\ge 0\,. \nonumber
\end{eqnarray}
Notice that $v^{(1)}$ decays to zero as $|r|\to\infty$ on a scale
of order $q^{-1}\sim{\cal O}(1)$.  Since $\kappa\ll 1$ 
we approximated $G$ in Eqn.~({\ref{vrst}) by
$G=(1+r\kappa)^{-1}\approx 1$.

For $n=2$ we obtain the stationary solution~\cite{HMRZ:97}
\begin{equation}
v^{(2)}(r,S,T)=\left[v^{(2)}_{\vert r=0} + {1\over 2}{v^{(1)}_{\vert r=0}}^2 
q^3 r\right]F(r)\,,\label{fun2}
\end{equation} 
For $n=3$, the stationary solution of (\ref{vn}) with (\ref{as_v3}) is
\begin{eqnarray}
v^{(3)}(r,S,T)=& 
-\frac{a_{00}}{q^2}+\left[v^{(3)}_{\vert r=0}+
\frac{a_{00}}{q^2}
+A_+ r - B_+ r^2 -C_+ r^3\right]e^{qr} & \qquad r\le 0\,, \\
v^{(3)}(r,S,T)=& 
-\frac{a_{00}}{q^2}+ \left[v^{(3)}_{\vert r=0}+
\frac{a_{00}}{q^2}
+ A_- r - B_- r^2 -C_-r^3\right]e^{-qr} & \qquad r\ge 0\,, 
\label{fun3}
\end{eqnarray}
where
\begin{eqnarray}
A_\pm &=&q^3v^{(1)}_{\vert r=0}v^{(2)}_{\vert r=0} \pm\left[{3\over 4q}V(0,S,T)+{1\over 2}q^5 {v^{(1)}_{\vert r=0}}^3
\right.-\left.{q\eta_1\over\eta_c}v^{(1)}_{\vert r=0} - qv^{(3)}_{\vert r=0}
+\frac{1-\delta^{-1}}{2q^2}\kappa_0\right]\,,  
\nonumber\\
B_\pm&=&{1\over 4}V(0,S,T) \pm q^4v^{(1)}_{\vert r=0}v^{(2)}_{\vert r=0}\,, 
\nonumber \\
C_\pm&=&\pm {1\over 6}q^7{v^{(1)}_{\vert r=0}}^3\,. \nonumber
\end{eqnarray} 
Application of the (no) jump condition $\bigl[v^{(3)}_r\bigr]_{r=0}=0$ leads
to 
\begin{equation}
{\partial v^{(1)}\over\partial T} = {\sqrt 2 \eta_1\over q\eta_c^2}v^{(1)}
- {3\over 4\eta_c^2} {v^{(1)}}^3 - {4\over 3}a_{00}
- {2(1-\delta^{-1})\over 3q}\kappa_0 + {\partial^2 v^{(1)}\over\partial 
S^2} - {\partial S\over\partial T}{\partial v^{(1)}\over\partial S}\ ,
\label{comp}
\end{equation}
where $v^{(1)}$ is evaluated at $r=0$. Using the expansion (\ref{vrStT}), the 
integral (\ref{st}), and transforming back to the fast variables $s,t$,
Eqn.~(\ref{comp}) becomes
\begin{equation}
{\partial v_f\over\partial t} = {\sqrt 2 (\eta_c-\eta)\over q\eta_c^2}v_f 
-{3\over 4\eta_c^2}v_f^3 - {4\over 3}a_0 \nonumber \\
-{2(1-\delta^{-1})\over 3q}\kappa
+ {\partial^2 v_f\over \partial s^2}-{\partial v_f\over\partial s}
\int_0^s \kappa C_n ds^\prime \ ,
\label{vfeqn}
\end{equation}
where $v_f(s,t)=v(0,s,t)$. 
Equation~(\ref{vfeqn}) coincides with the order parameter
equation~(\ref{C0}) once we make the 
following identifications: $C_0=-{3\over\eta\sqrt 2}v_f$, 
$\alpha={\eta\sqrt 2\over q\eta_c^2}$, $\alpha_c={\sqrt 2\over q\eta_c}$, 
$\beta=1/6$, $\gamma=\alpha_c(1-\delta^{-1})$,  and $\gamma_0=2\alpha_cqa_0$.

Algebraic $C_n-\kappa$ relations can be obtained from
Eqns.~(\ref{C0}) and (\ref{Cn}) for smooth weakly curved fronts assuming
$C_0$ follows adiabatically slow curvature variations:
\begin{equation}
C_n = C_0 - D\kappa\,,
\label{algebraic_Cn}
\end{equation}
where $C_0$ solves
\begin{equation}
(\alpha_c-\alpha)C_0 - \beta C_0^3 +\gamma\kappa + \gamma_0 = 0\,. 
\label{algebraic_C0}
\end{equation}
Such an assumption is valid away from the NIB bifurcation,
but the condition $C_0\sim c \ll 1$ used in deriving Eqn.~(\ref{C0})
no longer holds.
To test how Eqns.~(\ref{algebraic_Cn}) and (\ref{algebraic_C0}) 
perform away from the NIB bifurcation we compared them with
$C_n-\kappa$ relations 
obtained from the implicit equation 
\begin{equation}
C_n + D\kappa = \frac{3(C_n+\kappa)}{\sqrt 2 \eta q^2
  \sqrt{(C_n+\kappa)^2 +4q^2}} + \frac{3a_0}{\sqrt{2}\eta q^2}\,,
\label{implicit}
\end{equation}
derived in~\cite{HaMe:94c}.
Eqn.~(\ref{implicit}) is valid at any distance from the NIB bifurcation.
Figs.~\ref{fig:ck} show graphs of Eqns.~(\ref{algebraic_Cn}) and
(\ref{algebraic_C0}) (solid curves) and of solutions of
Eqn.~(\ref{implicit}) (dashed curves) close to and away from 
the NIB bifurcation.
The agreement between the two approaches
remains very good even
where $c > 1$ (Fig.~\ref{fig:ck}b).   
Thus the adiabatic elimination of $C_0$ away from the bifurcation
reproduces the algebraic $C_n-\kappa$ relations 
to a very good approximation.
\begin{figure}
\centering\includegraphics[width=6.5in]{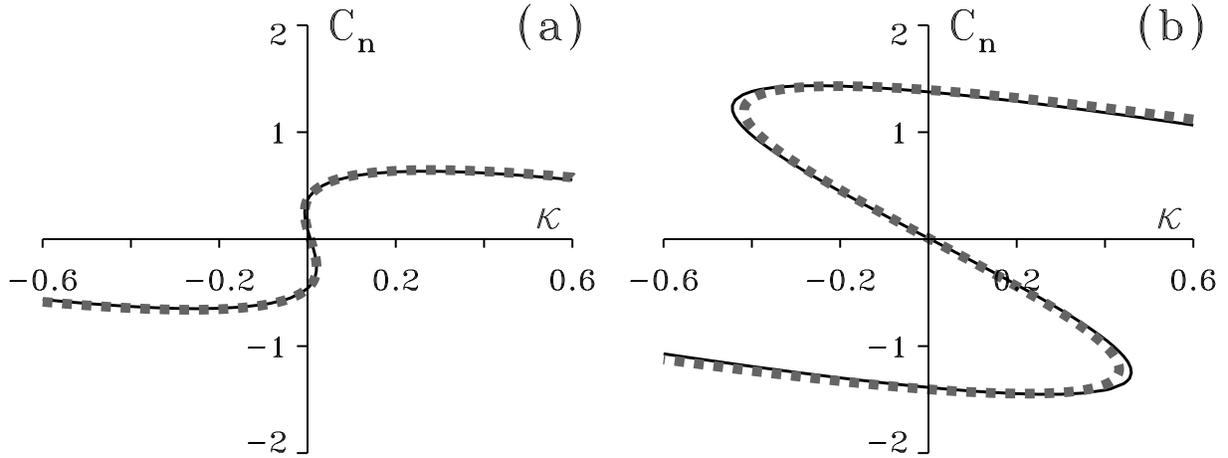}
\caption{
The $C_n-\kappa$ relations 
(\protect\ref{algebraic_Cn})-(\protect\ref{algebraic_C0})
derived from the kinematic equations
(solid curve) and solutions of 
the implicit relation (\protect\ref{implicit}) (dashed line). 
(a) Near the NIB bifurcation,
$\epsilon=0.0115$, both relations give the same result.  
(b) Farther from the bifurcation, $\epsilon=0.0105$,
the agreement is still good even though the kinematic
equations are derived for $c\ll 1$.
Parameters: $a_1=4$, $a_0=-0.0001$, $\delta=1.063$.
}
\label{fig:ck}
\end{figure}

%
%
\section{Numerical solutions of the kinematic equations}

We study two types of solutions to the kinematic equations: 
steadily rotating spiral waves (Section 4.1), and 
the nucleation of a spiral-wave pair by a transverse instability (Section 4.2). 
\subsection{Spiral waves}
Consider a ``front'' solution connecting the planar Bloch front,
$C_0=C_0^+$, $\kappa=0$, at $s=-\infty$ with the planar Bloch front,
$C_0=C_0^-$, $\kappa=0$, at $s=+\infty$, where $C_0^\pm=
\pm\sqrt{(\alpha_c-\alpha)/\beta}$, and we have assumed a symmetric model,
$a_0=0$ or $\gamma_0=0$. 
Fig.~\ref{fig:spirals}a shows such a solution obtained 
by numerically integrating~(\ref{C0})-(\ref{Cn}).
As demonstrated in Fig.~\ref{fig:spirals}b this front solution of the kinematic
equations~(\ref{C0})-(\ref{Cn}) represents a {\em spiral-wave} solution of
the FitzHugh-Nagumo model~(\ref{FHN}).  
Far away from the spiral core the leading front of the spiral
approaches a planar Bloch front pertaining to an up state invading
a down state ($C \to C_0^+$ as $s \to -\infty$).  
The trailing front approaches a planar
Bloch front pertaining to a down state invading an up state
($C \to C_0^-$ as $s \to +\infty$).  The spiral core is 
naturally captured as the interface separating these two Bloch fronts.
Cores of spiral waves in bistable and excitable media have
also been studied in 
Refs.~\cite{Fife:84,PeSu:91,Bernoff:91,Karma:92,Keener:92a,KLR:94} 
using steady state free boundary formulations with
linear $C_n-\kappa$ relations.
\begin{figure}
\centering\includegraphics[width=5.0in]{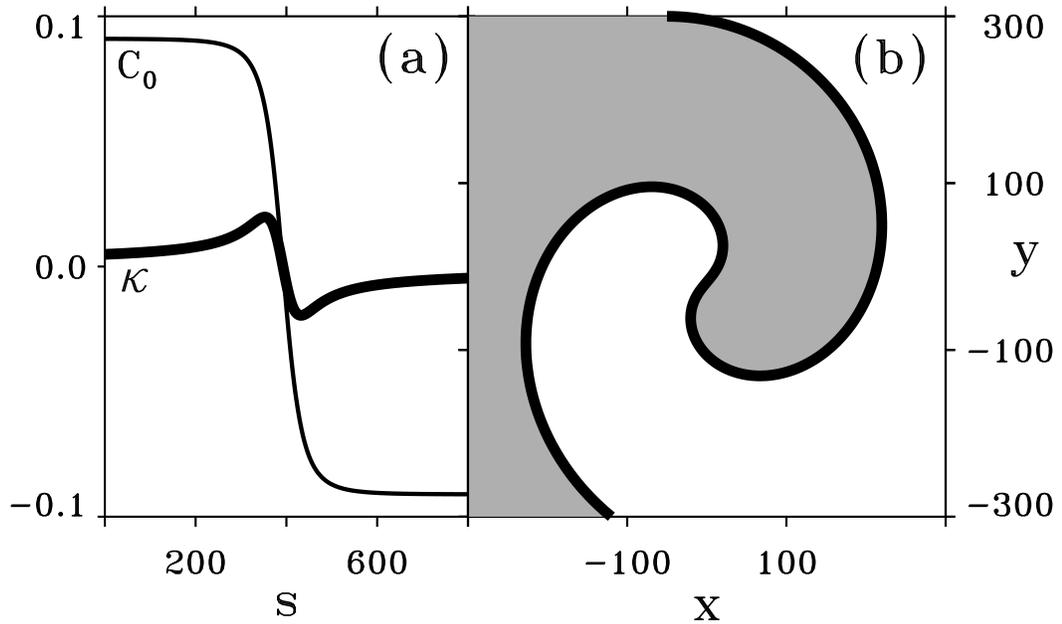}
\caption{
    A front solution to the kinematic equations
    (\protect\ref{C0})-(\protect\ref{Cn}).  (a) The order parameter
    $C_0$ and the curvature $\kappa$ along the arclength $s$.  (b) In
    the $x-y$ plane the front solution corresponds to a rotating
    spiral wave. The shaded (light) region corresponds to an up (down)
    state.  Parameters: $a_1=4.0$, $a_0=0$, $\epsilon=0.01234$,
    $\delta=1.0.$ 
}
\label{fig:spirals}
\end{figure}

Fig.~\ref{fig:spirala}a shows a similar front solution but for an
asymmetric system, $a_0\ne 0$ or
$\gamma_0\ne0$. Fig.~\ref{fig:spirala}b shows the corresponding spiral
wave. The spiral tip, defined as the point of zero curvature, is no
longer stationary, but rotates along a circle. A word of caution is
needed here, however. The kinematic equations do not take into account
interactions between front segments.  Including such interactions
would require changing the asymptotic conditions,
$v(\mp\infty,s,t)=v_\pm$ in Eqns.~(\ref{free}). As a result, away from the
tip (outside the frame in Fig.~\ref{fig:spirala}b),
the trailing front of the spiral meets the leading front
and the up state domain disappears.  

\begin{figure}
\centering\includegraphics[width=5.0in]{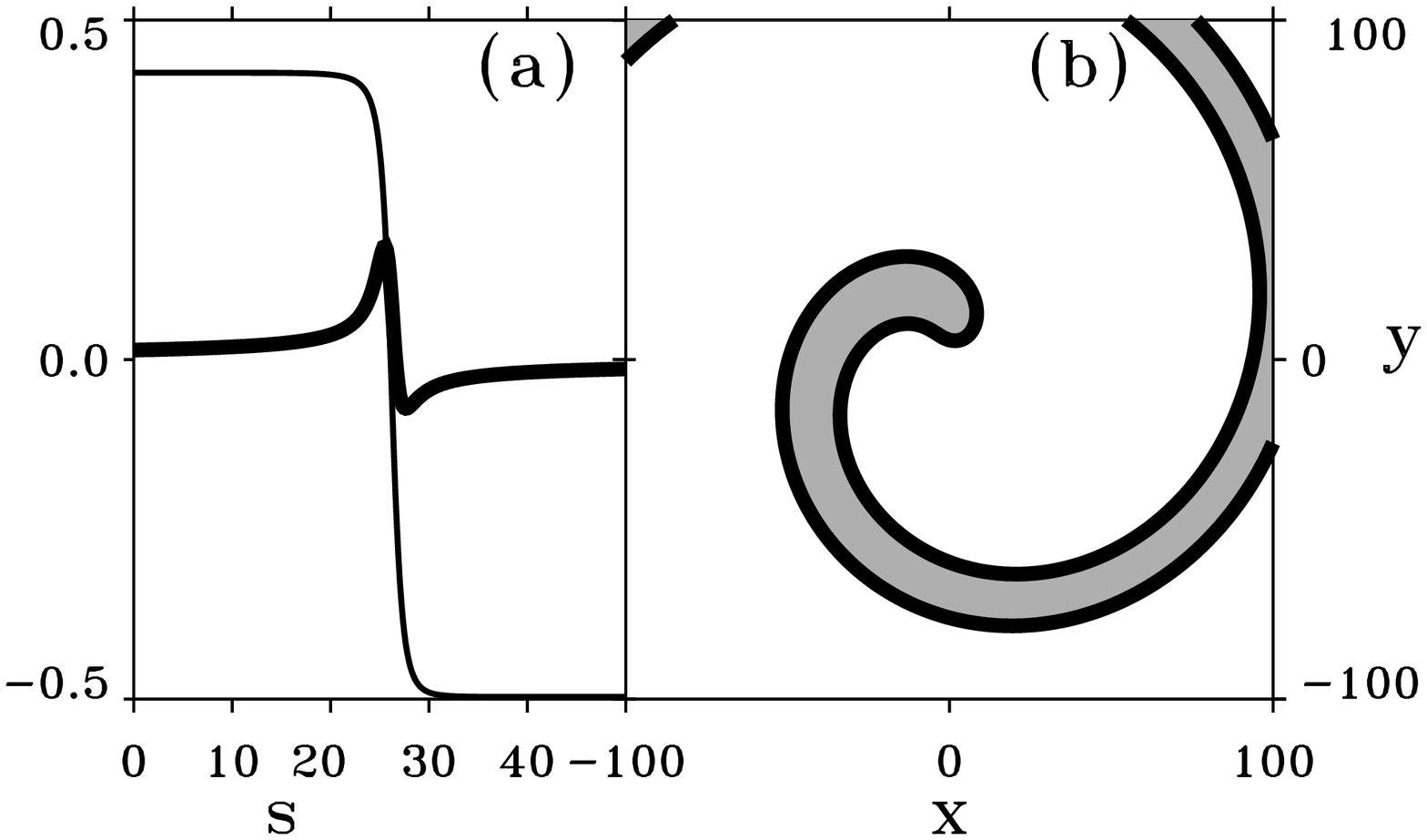}
\caption{
    A front solution to the kinematic equations for a nonsymmetric
    system ($\gamma_0 \ne 0$).  (a) The order parameter $C_0$ and the
    curvature $\kappa$ along the arclength $s$.  (b) In the $x-y$
    plane the front solution corresponds to a rotating spiral wave.
    The shaded (light) region
    corresponds to an up (down) state.  Parameters: $a_1=4.0$,
    $a_0=-0.0001$, $\epsilon=0.0115$, $\delta=1.0.$ 
}
\label{fig:spirala}
\end{figure}

\subsection{Spiral nucleation induced by a transverse instability}

Earlier numerical solutions of the FitzHugh-Nagumo model (\ref{FHN}) 
revealed that an instability of a planar front solution to transverse 
perturbations near a NIB bifurcation can induce spontaneous nucleation of 
spiral waves followed by domain splitting~\cite{HaMe:94b}. The kinematic 
equations (\ref{C0})-(\ref{Cn}) capture the transverse instability of 
the Ising front and, to linear order around the codimension 3 
point, $P3$, also the transverse instability boundary of the Bloch fronts. 
Explicit expressions for the transverse instability thresholds 
for a symmetric system ($\gamma_0=0$) can readily be obtained. 
Let 
\begin{eqnarray*}
C_0&=&C_0^0 + \bar C_0\exp(\sigma t+iQs)+c.c.\\ 
\kappa&=&\kappa^0 + \bar\kappa\exp(\sigma t+iQs)+c.c. 
\end{eqnarray*}
where
$(C_0^0,\kappa^0)=(0,0)$ for the Ising front and
$(C_0^0,\kappa^0)=(\pm\sqrt{(\alpha_c-\alpha)/\beta},0)$ for the Bloch
fronts.  Inserting these forms in~(\ref{C0})-(\ref{Cn}) gives the following
transverse instability lines, linearized around $\delta=1$:
\[
{\rm Ising:} ~~~\epsilon=\eta_c^2\delta\,, \qquad {\rm Bloch:}
~~\epsilon=\eta_c^2(3-2\delta) \,.
\]
These lines are displayed in Fig.~\ref{fig:lines} (thin lines). 
Fig.~\ref{fig:growth} shows typical growth 
rates of transverse perturbations of wavenumber $Q$ for Ising fronts 
(solid line) and Bloch fronts (dashed line) for the symmetric case.
Note that the first 
wavenumber to grow as the transverse instability lines are traversed 
is $Q=0$, consistent with our assumption of small curvature in the 
vicinities of (or below) these lines.
\begin{figure}
\centering\includegraphics[width=5.0in]{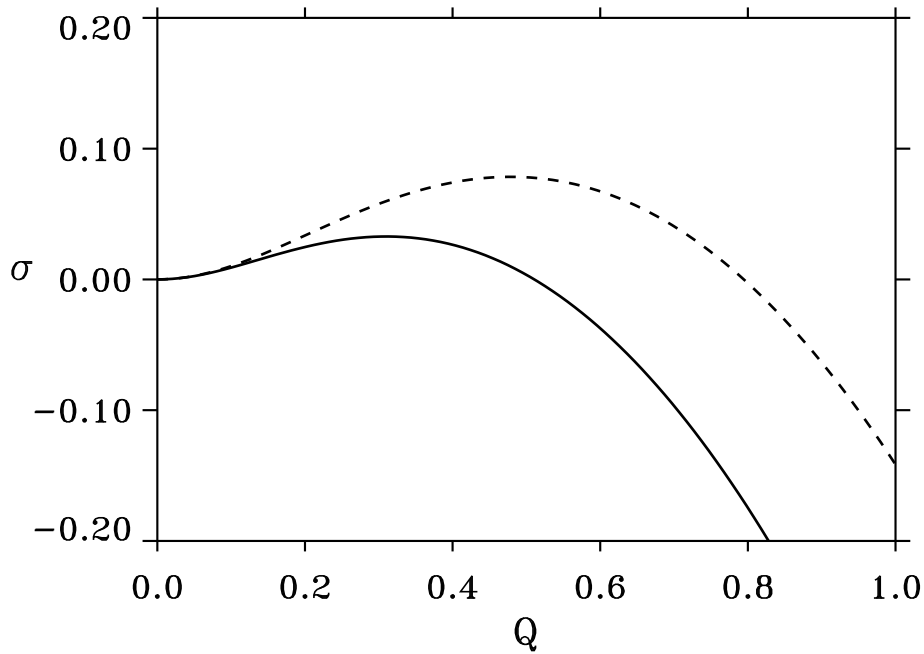}
\caption{ 
    Growth rates, $\sigma$, of transverse perturbations of 
    wavenumber $Q$ to uniform
    solutions of the kinematic 
    equations~(\protect\ref{C0})-(\protect\ref{Cn}).  
    Solid line: Bloch fronts, $\epsilon=0.011$, $\delta=1.08$.
    Dashed line: Ising fronts,  $\epsilon=0.012$, $\delta=1.2$.
    In both cases $a_1=4.0$, $a_0=0$.
}
\label{fig:growth}
\end{figure}

To test whether spiral wave nucleation induced by a transverse
instability is captured by the kinematic equations, we numerically
computed the time evolution of a planar front near the NIB bifurcation
and beyond the transverse instability
boundary. Figs.~\ref{fig:nucleation}a-d show four snapshots of this
time evolution.  The initial front pertains to an up state invading a
down state ($C_0>0$). The transverse instability causes a small dent
on the front to grow (Fig.~\ref{fig:nucleation}b). The negative
curvature then triggers the nucleation of a region along the arclength
where the propagation direction is reversed
($C_0<0$)~(Fig.~\ref{fig:nucleation}c). The pair of fronts in the
kinematic equations that bound this region correspond to a pair of
counter-rotating spiral waves in the FitzHugh-Nagumo equations
(Fig.~\ref{fig:nucleation}d).  Fig.~\ref{fig:nucleation}d demonstrates
the equivalence of spiral pair nucleation in the bistable medium to
``droplet'' nucleation in the kinematic equations. 
The $C_n-\kappa$ relation 
for the parameter values of Figs.~\ref{fig:nucleation} is shown
in Fig.~\ref{fig:ck}a.  Although the relation does not capture the nucleation
dynamics it does provide a heuristic explanation of the
nucleation process:  the negative curvature that develops at
the dent grows beyond the termination point of
the upper branch in Fig.~\ref{fig:ck}a
and a transition to the lower branch takes place.
This results in the reversal of propagation direction
and the nucleation of a spiral-wave pair.
The positive slopes
of the upper and lower branches at $\kappa=0$ indicate transverse
instabilities of the two planar Bloch front solutions.

The proximity to the front bifurcation is essential for 
spontaneous spiral-wave nucleation.
Farther from the bifurcation initial dents may grow due to the transverse
instability but not nucleate spiral waves.  
This is demonstrated in Figs.~\ref{fig:nonucleation} where
the same initial conditions as in Fig.~\ref{fig:nucleation}a
are chosen. The initial almost planar front
develops a dent but the dent retracts rather than nucleate a spiral-wave pair.
\begin{figure}
\centering\includegraphics[width=5.0in]{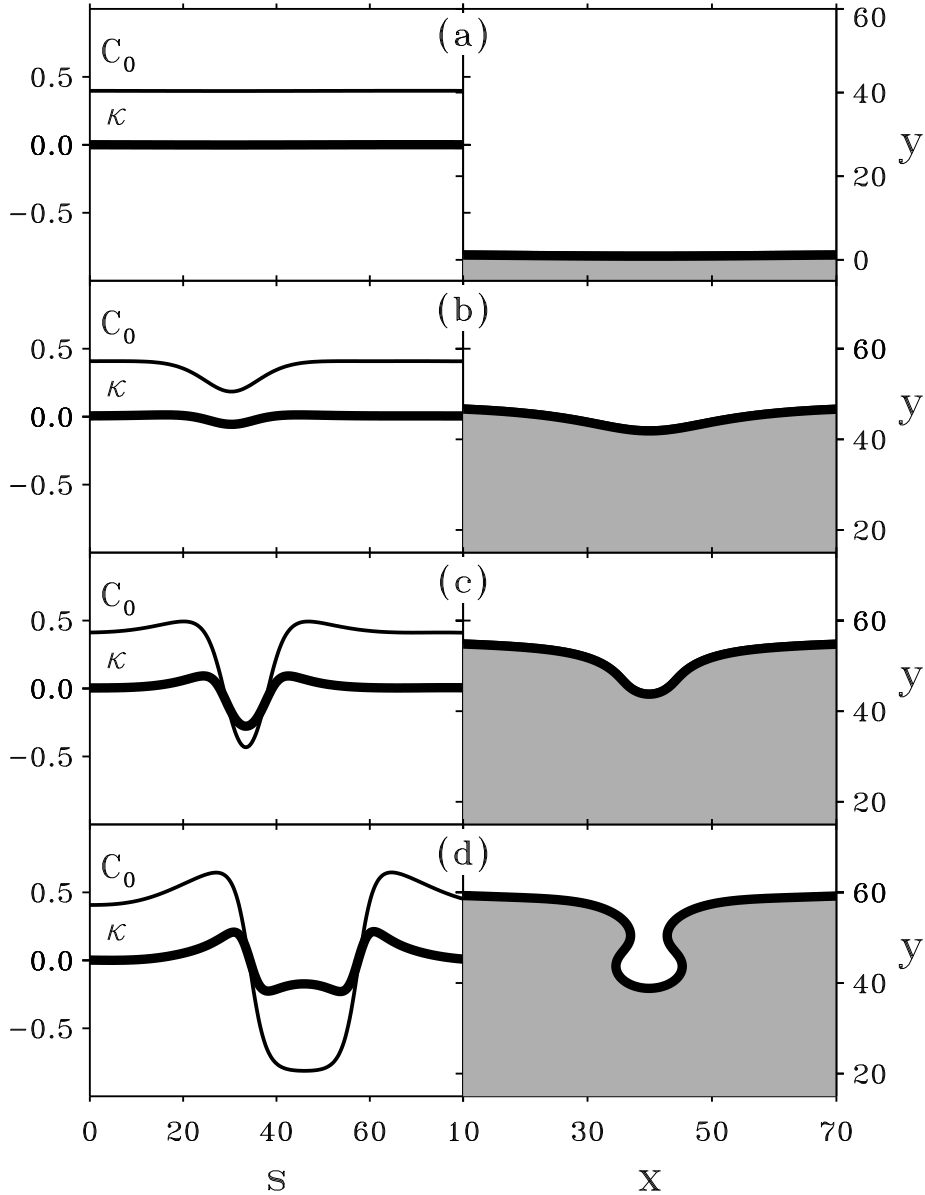}
\caption{
    A numerical solution of the kinematic equations
    (\protect\ref{C0})-(\protect\ref{Cn}) demonstrating
    the nucleation of a spiral-wave pair from an unstable
    propagating front.  
    Left column: the $C_0(s)$ and $\kappa(s)$ profiles.  
    Right column: the front line shape in the $x-y$ plane.  
    A small perturbation on the initially near planar front (a)
    grows (b) and nucleates a pair of spiral waves (c).  The front
    regions in the left frame of (d) each correspond to the core of
    the rotating spiral in the right frame.
    Parameters: $a_1=4.0$, $a_0=-0.0001$,
    $\epsilon=0.0115$, $\delta=1.063.$ (a)-(d) are at
    $t=0,116,136,142$.  
}
\label{fig:nucleation}
\end{figure}
\begin{figure}
\centering\includegraphics[width=5.0in]{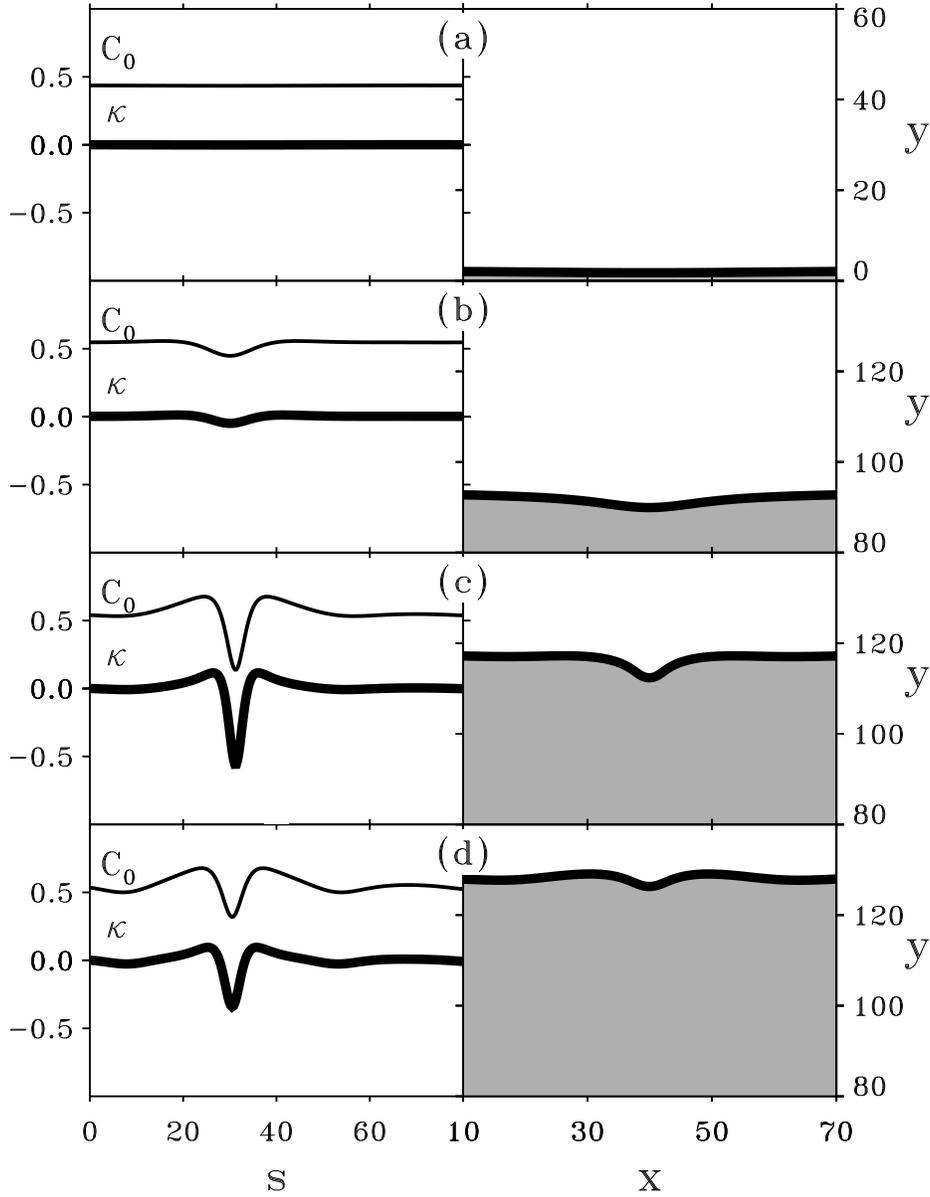}
\caption{
    A numerical solution of the kinematic equations
    (\protect\ref{C0})-(\protect\ref{Cn}) with
    the same initial conditions as in Fig.~\protect\ref{fig:nucleation}
    but farther away from the NIB bifurcation.
    Left column: the $C_0(s)$ and $\kappa(s)$ profiles.  
    Right column: the front line shape in the $x-y$ plane.  
    The transverse instability causes a small dent (a)
    to grow (b) but since the system is far enough from
    the NIB bifurcation, no spiral-wave nucleation occurs (c), and
    the dent contracts (d).
    Parameters: $a_1=4.0$, $a_0=-0.0001$,
    $\epsilon=0.0115$, $\delta=1.055.$ (a)-(d) are at
    $t=0,175,220,240$.  
}
\label{fig:nonucleation}
\end{figure}

%
%
\section{Conclusion}
We have developed a new set of kinematic equations for front dynamics
in nearly symmetric bistable media.  The equations are quantitatively
valid for slowly propagating, weakly curved fronts. They 
describe the growth of transverse perturbations, the core structure
of spiral waves, and the dynamics
of spiral-wave nucleation. Away from the NIB bifurcation, where the
front velocity is no longer a slow degree of freedom, the algebraic
$C_n$-$\kappa$ relation is recovered.

The process of spiral-wave nucleation involves local transitions
between the counterpropagating Bloch fronts. In the
present work these transitions were driven by curvature
perturbations. Front transitions and spiral nucleation can also be
driven by other intrinsic perturbations, such as nonlocal front
interactions or 
interactions with boundaries~\cite{Haim:96}.  Such
interactions, however, have not been included
in the present derivation. They are excluded by the choice of the
boundary conditions in Eqns.~(\ref{free}).  Front interactions are not
important for the study of symmetric spirals (Figs.~\ref{fig:spirals})
or the initial nucleation of a spiral-wave pair from a planar front
(Figs.~\ref{fig:nucleation}).  They do, however, affect nonsymmetric
spiral-waves ($a_0\ne0$) and might play an important role in the
meander instability of a spiral tip~\cite{Winfree:91,Barkley:94,KK:97}.
Front interactions are also essential for the formation of
labyrinthine patterns in the Ising regime~\cite{GMP:96}.

The kinematic equations generalize an earlier approach based on a 
geometric equation for curvature supplemented by a linear normal
velocity - curvature relation~\cite{Zykov:87,Mikhailov:90,Meron:92,MDZy:94}.  
In that case the speed of a planar front is taken to be a constant, determined
by the parameters of the system, and no distinction is made between
the two types of Bloch fronts. This approach 
has been applied to traveling waves in excitable media
modeling a pulse stripe as a single curve~\cite{BrTy:96,Brazhnik:96,PSG:96}.
It has also been applied to spiral-wave dynamics 
with phenomenological assumptions to describe the motion of the spiral 
tip, the free end of the 
curve~\cite{Zykov:87,Mikhailov:90,Meron:92,MDZy:94,MePe:88}. 
No phenomenological assumptions
are needed in applying the generalized kinematic equations as they naturally
describe the core (or tip) structure of a  spiral-wave. 

Finally, we note that
the two-dimensional spiral-wave nucleation problem in the original
bistable medium is reduced to a one-dimensional
``droplet'' nucleation problem in the kinematic 
equations~\cite{Fife:79,GuDr:83}.
This may simplify the evaluation of the critical curvature perturbation
required to nucleate spiral waves.

\ack
We wish to thank Paul Fife for many interesting discussions. This research
was supported in part by grant No.~95-00112 from the US-Israel Binational 
Science Foundation (BSF).

\bibliography{reaction}

\end{document}